\documentstyle[sprocl,psfig]{article}
\def\be{\begin{equation}}
\def\ee{\end{equation}}
\def\bea{\begin{eqnarray}}
\def\eea{\end{eqnarray}}

\begin{document}

\title{INSTANTONS AND NUCLEON MAGNETISM }
\author{H. FORKEL}
\address{Institut f{\"u}r Theoretische Physik, Universit{\"a}t Heidelberg, \\ 
D-69120 Heidelberg, Germany}
\maketitle 
%%%%%%%%%%%%%%%%%%%%%%%%%%%%%%%%%%%%%%%%%%%%%%%%%%%

\abstracts{ We construct improved QCD sum rules for the nucleon magnetic 
moments by implementing direct-instanton contributions to the operator 
product expansion of the nucleon correlator in a magnetic background field. 
The instanton contributions are found to affect only those sum 
rules which had previously been considered unstable. The new sum rules 
show a high degree of stability and reproduce the experimental values 
of the nucleon magnetic 
moments for values of the magnetic quark condensate susceptibility which 
are consistent with other estimates. (Invited talk given at ``Hadron 
Physics 2000'', Caraguatatuba, S\~ao Paulo, Brazil (April 10-15,
2000). }

Over the last years, an increasingly multifaceted picture of the role of QCD
instantons \cite{bel75} in hadron physics has begun to emerge from hadron
models with more or less instanton-induced structure \cite{hor78},
instanton-liquid vacuum models\cite{sch98}, a generalized operator product
expansion including instanton contributions (IOPE) \cite{for00} and, most
recently, lattice calculations \cite{lat}. 

One of the major benefits of sharpening this picture by identifying specific
hadronic instanton effects will be to end  the puzzling elusiveness of 
explicit glue in hadrons at low energies. Indeed, in most successful models
of the classical mesons and baryons, gluons play at best a marginal role. 
The situation is further obfuscated by the fact that gluons do not directly
couple to electroweak probes. Since much of what we know about the classical
hadrons derives from their response to such probes, it is particularly
desirable to pin down traces of nonperturbative glue in electromagnetic
hadron observables.

In this talk, I will report on an investigation \cite{aw99} which takes a
step towards uncovering the hidden role of glue in the response to
electromagnetic fields by studying the impact of instantons on the nucleon
magnetic moments in the recently developed IOPE sum-rule approach \cite
{for00,FB}. While in general the range of applicability of the IOPE (as 
a short-distance expansion) is more limited than that of instanton vacuum
models or lattice calculations, it profits from the transparency of an
analytical (yet largely model-independent) method and takes, in contrast to
instanton models, all long-wavelength vacuum fields and also perturbative
fluctuations into account. 

In order to study the nucleon's magnetic moments in the framework of the 
IOPE, we start from a correlation function which contains the response of
the nucleon to the electromagnetic current at short, spacelike distances. 
(The information on the magnetic moments is subsequently extracted by
dispersion techniques \cite{for00,svz79}.) At first sight, the natural
candidate for such a correlation function would be 
\begin{equation}
\Pi _{\mu }\left( p,q\right) =-\int d^{4}xe^{ipx}\int
d^{4}ye^{iqy}\left\langle 0\left| T\eta \left( x\right) J_{\mu } 
\left( y\right) \bar{\eta}\left( 0\right) \right| 0\right\rangle 
\label{corr0}
\end{equation}
where the nucleon interpolating fields $\eta (x)$ (with either proton or
neutron quantum numbers) are composite operators of massless up and down
quark fields \cite{I}, 
\begin{equation}
\eta _{p}(x)=\epsilon ^{abc} [u^{a^{T}}(x)C\gamma _{\alpha }u^{b}(x)]
\gamma_{5}\gamma ^{\alpha }d^{c}(x),\qquad \eta_{n}
=\eta_{p}(u\leftrightarrow d),  \label{interpol} \end{equation}
and $J_{\mu }$ is the electromagnetic current. However,
obtaining the nucleon magnetic moments requires taking the $q\rightarrow 0$
limit in which the matrix element of (\ref{corr0}) is probed mainly 
in the large-distance region $\left| y\right| ,\left|
x-y\right| \gg \Lambda ^{-1}$ where the operator product expansion breaks
down\footnote{
The OPE\ of this matrix element can be used, however, to obtain the nucleon
form factors at intermediate momentum transfers $Q\sim 1$ GeV \cite{bel93}$.$
}.

Fortunately, there exists an alternative approach, due to Ioffe and Smilga
\cite {IS} and Balitsky and Yung \cite{BY}, which can be directly applied to
the calculation of the magnetic moments. The underlying idea is to treat the
long-wavelength photons at the same level as the soft QCD vacuum fields
(which make up the condensates), i.e. to account for them in the operators
of the (I)OPE (see below). This amounts to considering the electromagnetic
field as a constant, classical background field (which, incidentally, is
reminiscent of the experimental setup used in actual measurements of the 
moments). In other words, the electromagnetic probe does not
appear anymore as an external field but rather as part of the 
``magnetized'' vacuum state.

We are thus led to consider the nucleon correlation function 
\begin{eqnarray}
\Pi (p) &=&i\int d^{4}x\,e^{ipx}{\langle 0|T\eta (x)\bar{\eta}(0)|0\rangle }
_{F}  \label{corr1} \\
&=&\Pi _{0}(p)+\sqrt{4\pi \alpha }\Pi _{\mu \nu }(p)F^{\mu \nu }+O\left(
F^{2}\right)   \label{corr4}
\end{eqnarray}
in the background of a constant electromagnetic field $F_{\mu \nu }$. We
have not written down higher orders of $F$ since the information on the
magnetic moments is part of the linear response 
\begin{equation}
\Pi _{\mu \nu }(p)=(\rlap/p\sigma _{\mu \nu }+\sigma _{\mu \nu }\rlap/
p)\,\Pi _{1}(p^{2})+i(\gamma _{\mu }p_{\nu }-\gamma _{\nu }p_{\mu })\rlap/
p\,\,\Pi _{2}(p^{2})+\sigma _{\mu \nu }\,\Pi _{3}(p^{2})  \label{lresp}
\end{equation}
to the (arbitrarily weak) external field. Above we have given the general
decomposition of $\Pi _{\mu \nu }(p)$ in terms of three independent Lorentz
and spinor structures which are associated with one chirally-even ($\Pi _{1}$)
and two chirally-odd invariant amplitudes ($\Pi _{2}$ and $\Pi _{3}$). The
relation to the magnetic moments can be made explicit by writing 
\begin{equation}
\Pi (p)=i\int d^{4}x\,e^{ipx}{\langle 0|T\eta (x)\bar{\eta}(0)e}^{-i\int
d^{4}yA^{\mu }J_{\mu }}{|0\rangle }  \label{corr3}
\end{equation}
where $J_{\mu }={\bar{q}\gamma }_{\mu }{Qq}$ is the
electromagnetic current with quark charge matrix $Q$ and $A_{\mu }$
is the external vector potential. In
fixed-point gauge, and specializing to a constant magnetic background field
 \begin{equation}
B_{i}=-\frac{1}{2}\varepsilon_{ijk} F_{jk}, \hspace{1.2cm}
\left(E_{i}=F_{i0}=0\right) 
\end{equation}
with the potential 
\begin{equation}
A_{\mu }\left( y\right) =-\frac{1}{2}F_{\mu \nu }y^{\nu },
\end{equation}
the above exponential can be rewritten as 
\begin{equation}
\int d^{4}yA^{\mu }J_{\mu }=\frac{1}{2}F_{ij}\int
d^{4}yy_{j}J_{i}=\int dt\vec{\mu}\cdot \vec{B}
\end{equation}
where $\vec{\mu}$ is the magnetic-moment operator 
\begin{equation}
\vec{\mu}=\frac{1}{2}\int d^{3}x\left( \vec{x}\times \vec{J}
\right) .
\end{equation}
Expanding (\ref{corr3}) to first order in $F$ and inserting nucleon
intermediate states, we thus obtain 
\begin{equation}
\Pi _{\mu \nu }(p)F^{\mu \nu }=-\varepsilon _{ijk}\Pi _{ij}(p)B_{k}\sim {
\langle N}\left( p\right) {|}\vec{\mu}{|N}\left( p\right) {\rangle }\cdot 
\vec{B}  \label{polecontrib}
\end{equation}
in terms of the nucleon magnetic moments, as anticipated.

In order to write down sum rules for ${\langle N}\left( p\right) {|}\vec{\mu}
{|N}\left( p\right) {\rangle }$, we need a QCD description of the correlator
(\ref{corr1}) at momenta $s=-p^{2}\simeq 1{\rm GeV}^{2}$, i.e. at distances 
$x\sim 0.2\,{\rm fm}$. Such a description is provided by the nonperturbative
IOPE  which factorizes (\ref{corr1}) into contributions from soft ($
k_{i}<\nu $) and hard ($k_{i}\geq \nu $) field modes (with momenta $k_i$),
where $\nu \sim 0.5\,{\rm GeV}$ is the operator renomalization scale. The
IOPE is generated by splitting each diagram contributing to (\ref{lresp}) in
all possible ways into a hard and a soft subgraph. The hard subgraphs, with
the integration range of each internal momentum restricted\footnote{
In practice, this restriction is often unnecessary.} to be larger than $\nu$
, give rise to the Wilson coefficients and receive, beyond the standard
perturbative contributions \cite{IS}, direct instanton
contributions which we will evaluate below. The soft subgraphs yield 
hadron-channel independent condensates, i.e. vacuum expectation values of
local, composite QCD operators renormalized at $\nu$. 

The new feature brought in by the magnetic background field is that the
``magnetized'' vacuum state $\left| 0\right\rangle _{F}$ singles out a
preferred direction (that of the field strength $\vec{B}$) and therefore
ceases to be a Lorentz scalar. Hence, several Lorentz-covariant operators 
of the OPE (considered here up to dimension eight) aquire finite expectation
values, the $F$-induced condensates 
\begin{eqnarray}
\langle 0|\bar{q}\sigma _{\mu \nu }q|0\rangle _{F} &=&\sqrt{4\pi \alpha }
\chi F_{\mu \nu }\langle 0|\bar{q}q|0\rangle ,  \label{magncond1} \\
g\langle 0|\bar{q}G_{\mu \nu }q|0\rangle _{F} &=&\sqrt{4\pi \alpha }\kappa
F_{\mu \nu }\langle 0|\bar{q}q|0\rangle , \\
g\langle 0|\bar{q}\gamma _{5}\tilde{G}_{\mu \nu }q|0\rangle _{F} &=&
\frac{i}{2}\sqrt{4\pi \alpha }\xi F_{\mu \nu }\langle 0|\bar{q}q|0\rangle .
\end{eqnarray}
($G_{\mu \nu }=\frac{1}{2}\lambda _{a}G_{\mu \nu }^{a}$, $\tilde{G}_{\mu \nu
}=\frac{1}{2}\varepsilon _{\mu \nu \rho \sigma }G_{\rho \sigma }$ with 
$\varepsilon _{0123}=-1$.) The parameters $\chi ,\kappa $, and $\xi$
quantify the vacuum response to weak electromagnetic fields and thus play
the role of generalized susceptibilities. The magnetic susceptibility of the
quark condensate, $\chi $, for example, originates from the induced spin
alignment of quark-antiquark pairs in the vacuum. The  corresponding vacuum
expectation value is the lowest-dimensional induced condensate and
will therefore play a leading role in the magnetic moment sum rules.

Altogether, the IOPE of the invariant amplitudes $\Pi _{i}(p)$ in Eq. (\ref
{lresp}) can thus be written as 
\begin{equation}
\Pi _{i}(p^{2})=\sum_{n}C_{i,n}\left( p^{2};\nu \right) \langle 0|
{\cal O}_{n}\left[ \nu \right] |0\rangle _{F}  \label{iope}
\end{equation}
where the ${\cal O}_{n}$ are local, composite QCD operators with the
appropriate quantum numbers (the multi-index $n$ can contain Lorentz
indices) and dimension $d\leq 8$\footnote{
Contributions from higher-dimensional operators are strongly suppressed, see
below.}. The $C_{n}$ are the Wilson coefficients. It might be worth
reiterating that the instanton contributions to (\ref{iope}) factorize into
those from soft modes (contained in the condensates $\langle 0|
{\cal O}_{n}|0\rangle $) and those due to hard modes which add nonperturbative
structure to the $C_{n}$. The latter can be of substantial size since
$\bar{\rho}<\nu ^{-1}\simeq 0.4\,{\rm fm}$ and have to be calculated explicitly.
(Nevertheless, they were neglected in the usual, perturbative treatment.) 

The calculation of the IOPE coefficients is described in Ref. \cite
{for00,aw99} to which we refer for details. Similar to the
phenomenologically known values of the lowest-dimensional condensates,
the bulk properties of the instanton size distribution are generated by
long-distance vacuum dynamics and thus taken as input. We will use the
standard ``instanton liquid'' values \cite{sch98} $\bar{\rho}\simeq
\frac{1}{3}\,{\rm fm}$ for the average instanton size and $\bar{R}\simeq 1{\rm fm}$
for the average separation between neighboring (anti)instantons. Since the
instanton size distribution $n(\rho )$ is rather sharply peaked around 
$\bar{\rho}$, we can thus approximately set $n(\rho )=\bar{n}\delta (\rho
-\bar{\rho})$ with $\bar{n}\simeq $ 0.5 fm$^{-4}$. Moreover,
$\bar{\rho}^{-1}\gg \Lambda _{QCD} $ implies that the hard instanton
contributions can be calculated semiclassically. Since, finally,
multi-instanton correlations are negligible at the relevant distances $x\sim
0.2\,{\rm fm}\ll \bar{R}$, this amounts to evaluating the hard subgraphs in
the background of an (anti-) instanton field and subsequently averaging over
the instanton-parameter distribution. The main contributions typically arise
from the quark zero modes in the instanton field (which aquire an effective
mass \cite{shi80}  $\bar{m}(\rho )=-\frac{2}{3}\pi^{2}\rho^{2}\langle
\bar{q}q\rangle $ due to the interactions with ambient vacuum fields)
 and can be treated exactly, while
the continuum modes are approximated by plane waves.

More specifically, the leading instanton contributions to the correlator
(\ref{corr1}) can be traced to the graph in which two of the quarks propagate
in zero modes while the third interacts with the background field through the
magnetized quark condensate \footnote{Background-field induced transitions
between zero and continuum modes (which can be dominant in other correlators
\cite{for95}) are absent here.}. This contribution is purely nonperturbative
and difficult to account for, e.g., in quark-based hadron models \cite{for97}.
After averaging over the instanton size distribution, it reads  
\begin{eqnarray}
{\langle 0|T\eta _{p}(x)\bar{\eta}_{p}(0)|0\rangle }_{F,inst} &=&\Pi
_{0}^{inst}(x)-\frac{2^{3}e_{u}}{3\pi ^{4}}\frac{\bar{\rho}^{4}}{\bar{m}^{2}}
\times   \nonumber \\
&&\langle \bar{q}\sigma _{\mu \nu }q\rangle _{F}\,\sigma _{\mu \nu }\int
d^{4}x_{0}\frac{1}{(r^{2}+\bar{\rho}^{2})^{3}(x_{0}^{2}+\bar{\rho}^{2})^{3}}
\label{instinx}
\end{eqnarray}
for the  proton. $\Pi_{0}^{inst}$ contains the instanton contributions for 
$F=0$, calculated in \cite{FB}, and $r=x-x_{0}$, where $x_{0}$ specifies the
center of the instanton. The corresponding neutron correlator is
obtained by replacing $e_{u}$ with $e_{d}$. Above, we have used the
self-consistency condition \cite{CDG} 
\begin{equation}
\langle \bar{q}q\rangle =-2\int
d\rho \,\frac{n(\rho )}{\bar{m}(\rho )}=-2\,
\frac{\bar{n}}{\bar{m}(\bar{\rho})}
\end{equation}
to eliminate the $\bar{n}$ dependence
from (\ref{instinx}). For later use in the sum rules, we now calculate the
Fourier and Borel transform\footnote{ For the Borel transform we follow the
convention of Ref. \cite{I}.} of (\ref {instinx}), from which we obtain   
\begin{equation} \widehat{\Pi }_{3}(M^{2})=\frac{e_{u}}{128\pi ^{4}}a\chi
\bar{\rho}^{2}M^{6}I(z^{2})
\label{incontr2}
\end{equation}
where $M$ denotes the Borel mass parameter and $z= M\bar{\rho}$, $%
a= -(2\pi )^{2}\langle \bar{q}q\rangle $. The zero-mode loop gives rise
to the integral 
\begin{equation}
I(z^{2})=\int_{0}^{1}\frac{d\alpha }{\alpha ^{2}(1-\alpha )^{2}}e^{-
\frac{z^{2}}{4\alpha (1-\alpha )}}=4e^{-\frac{z^{2}}{2}}\left[ K_{0}
\left( \frac{z^{2}}{2}\right) +K_{1}\left( \frac{z^{2}}{2}\right) \right]
\label{instint}
\end{equation}
and exhibits the unique exponential Borel-mass dependence which is
characteristic for instanton contributions \cite{FB}.

A first important lesson of Eq. (\ref{incontr2}) is that, to leading order,
direct instantons contribute almost exclusively\footnote{
The small direct--instanton contributions to $\Pi _{2}$ have no appreciable
impact on the sum rules and are neglected.} to the chirally-odd amplitude
$\Pi_{3}$. This is crucial because in the analysis of Ref. \cite{IS}, were
this contribution was not accounted for, exactly the $\Pi_{3}$ sum-rule
failed to show a fiducial stability region \cite{ISP}. The  previous neglect
of the instanton contributions thus offers a potential explanation for
this instability. An additional hint in this direction comes from two other
chirally-odd nucleon sum rules \cite{FB,for97} where stabilization due
to instantons was indeed found to take place\footnote{ A previously
contemplated connection between the instability of the original  $\Pi_{3}$
sum rule and the perturbative infrared singularities of the ``pragmatic'' OPE
seems, in view of newer results \cite{WPC}, unlikely \cite{aw99}.}. 

We are now ready to proceed to the quantitative analysis of the IOPE sum
rule for $\Pi_{3}$. The latter results from equating the Borel transform of
the standard OPE of Ref. \cite{IS} and the instanton contributions (\ref
{incontr2}) to the Borel-transformed double dispersion relation for the
correlator (\ref{corr1}), with a spectral function parametrized in terms of
the nucleon pole contribution (cf. Eq. (\ref{polecontrib})) and a continuum
based on local duality. Including the infrared-divergent term encountered in
Ref. \cite{IS}, duly truncated at the OPE renormalization point, the IOPE
sum rule (for the proton) reads 
\begin{eqnarray}
&&aM^{2}\left\{ \left[ e_{u}-\frac{1}{6}e_{d}(1+4\kappa +2\xi )\right]
E_{1}(M)\right.   \nonumber \\
&&+\frac{1}{6}e_{u}\frac{m_{0}^{2}}{M^{2}}\left[ \ln {\frac{M^{2}}{\nu^{2}}}
-\gamma _{EM}\right] L^{-\frac{4}{9}}+\frac{1}{6}e_{d}M^{2}\chi
E_{2}(M) L^{-\frac{16}{27}}  \nonumber \\
&&\left. -\frac{1}{8}e_{u}\chi \rho _{c}^{2}M^{4}I(z^{2})L^{
-\frac{16}{27}}\right\}   \nonumber \\
&=&\frac{1}{4}\tilde{\lambda}_{N}^{2}me^{-\frac{m^{2}}{M^{2}}}\left[
\frac{\mu _{p}}{M^{2}}-\frac{\mu _{p}^{a}}{2m^{2}}+A_{p}\right],
\label{sumrule}
\end{eqnarray}
where $m$ is the nucleon mass, $\mu _{p}$ ($\mu _{p}^{a}$) the proton's
(anomalous) magnetic moment, $W$ the continuum threshold, and $\lambda _{N}$
the coupling of the current (\ref{interpol}) to the nucleon state, $\langle
0|\eta |N\rangle =\lambda _{N}u$. For the mixed quark condensate we use the
standard parametrization $\langle \bar{q}\sigma _{\mu \nu }G^{\mu \nu
}q\rangle =-m_{0}^{2}\langle \bar{q}q\rangle $ with $m_{0}^{2}=0.8\,
{\rm GeV}^{2}$, and $\gamma _{EM}\simeq 0.577$ is the Euler-Mascheroni
constant. The additional parameters $A_{p,n}$ determine the strength of
electromagnetically induced transitions between the nucleon and its excited
states. The sum rule for the neutron is obtained from (\ref{sumrule}) by
interchanging $e_{u}$ and $e_{d}$ and by replacing $\mu _{p},\mu
_{p}^{a}\rightarrow \mu _{n}$ and $A_{p}\rightarrow A_{n}$. We have also
defined $\tilde{\lambda}_{N}^{2}=32\pi ^{4}\lambda _{N}^{2}$, $L=\ln
(M/\Lambda )/\ln (\mu /\Lambda )$ ($\Lambda =0.1\,{\rm GeV}$), and
transferred, using the standard expressions 
\begin{equation}
E_{n}(M)=1-e^{-\frac{W^{2}}{M^{2}}}\left[ 1+\sum_{1}^{n}\frac{1}{j}\left( 
\frac{W^{2}}{M^{2}}\right) ^{j}\right] ,
\end{equation}
the continuum contributions to the IOPE-side of the sum rules. The
appropriate form of these contributions has recently been clarified in Ref. 
\cite{iof95}.

The quantitative analysis of the IOPE\ sum rule (\ref{sumrule}) is based on
the minimization of the relative deviation  between its two sides. Since the
fiducial domain, i.e. the Borel-mass region in which the IOPE truncation is
justified while the nucleon pole still dominates over the continuum, is (as
usual) not large enough to determine all unknown parameters from a combined
fit, we follow the procedure of Ref. \cite{IS} and obtain the coupling
$\tilde{\lambda}^{2}=2.93\,{\rm GeV}^{6}$ and the continuum threshold
$W=1.66\,{\rm GeV}$ by fitting the instanton-improved nucleon mass sum rule
of Ref. \cite{FB} to the experimental nucleon mass. The values of the two
susceptibilities $\kappa =-0.34\pm 0.1$ and $\xi =-0.74\pm 0.2$ were
estimated in independent work by Kogan and Wyler \cite{KW}. This enables us
to fit both sides of the sum rules (\ref{sumrule}) by varying $\chi $
and $A_{p}$ (or $A_{n}$, respectively) while keeping the magnetic
moments fixed at their experimental values.

The fits are performed in the fiducial Borel mass domain $0.8\,{\rm GeV}\leq
M\leq 1.15\,{\rm GeV,}$ where the highest-dimensional operators contribute
at most 10 \% to the IOPE and the continuum contribution does not
exceed 50 \%. This fiducial domain is larger, incidentally, than
that of the sum rules based on $\Pi_{1}$ and $\Pi_{2}$.

\begin{figure}[tbp]
\psfig{figure=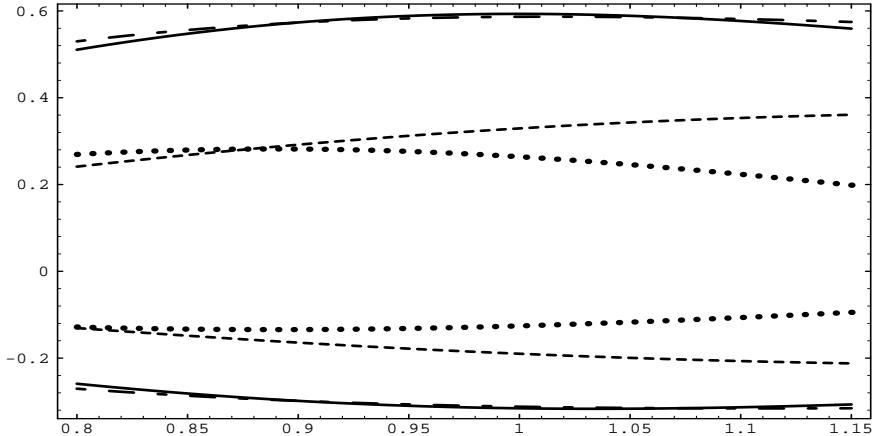,height=2.7in,width=4.5in,bbllx=72bp,bblly=230bp,%
bburx=539pt,bbury=570pt}
\caption{The OPE (dashed line) and direct instanton (dotted line)
contributions to the new $\protect\sigma_{\protect\mu\protect\nu}$ sum rules
for the proton (positive range) and neutron. Their sum (dot-dashed line) is
compared to the RHS (solid line).}
\label{fig1}
\end{figure}

\begin{figure}[tbp]
\psfig{figure=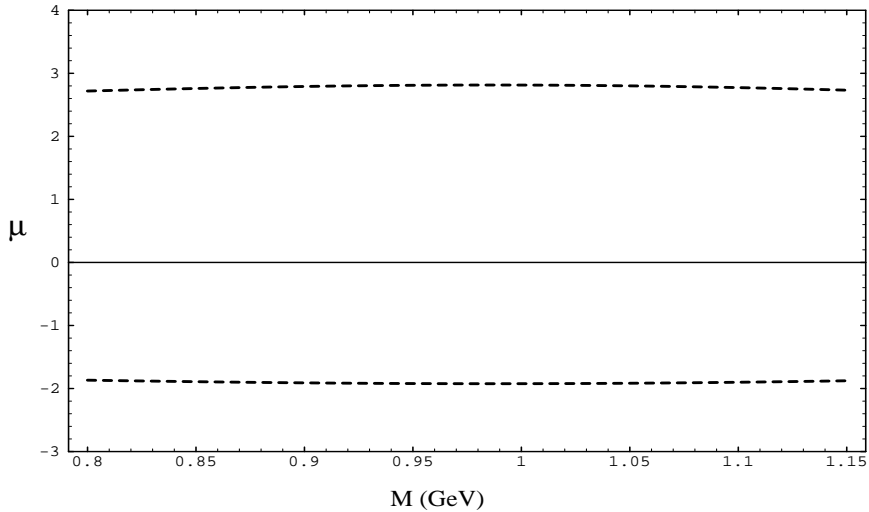,height=2.7in,width=4.5in,angle=-90}
\caption{The Borel mass dependence of the magnetic moments of the proton
(upper) and neutron calculated from the optimal fit of LHS and RHS.}
\label{fig2}
\end{figure}

Figure \ref{fig1} shows, both for the proton and the neutron sum rules, the
direct-instanton contributions, the remaining OPE including the continuum
contributions, their sum (which makes up the left-hand side of Eq. (\ref
{sumrule})), and the right-hand sides. The fit quality of both the proton
and neutron sum rules is quite impressive. Figure \ref{fig1} shows,
furthermore, that the direct-instanton contributions can reach 
the magnitude of the remaining terms in the OPE, which explains why their
previous neglect had a detrimental impact on the sum-rule stability. In Fig.
\ref{fig2} the optimized sum rules are solved for $\mu_{N}$ and plotted as a
function of the Borel mass. The resulting functions $\mu_{p,n}(M)$ therefore
specify the value of the magnetic moment which is required to make both sides
of the sum rule (\ref{sumrule}) match exactly at a given value of $M$. As a
consequence of the high fit quality, $\mu (M)$ is practically 
$M$-independent. The numerical results are $\chi \simeq -4.96\,{\rm GeV}^{-2}$, 
$A_{p}\simeq 0.28{\rm GeV}^{2}$ for the proton and $\chi \simeq -4.73\,{\rm
GeV}^{-2}$, $A_{n}\simeq -0.27{\rm GeV}^{-2}$ for the neutron sum rule. The
values of the quark condensate susceptibility $\chi $ (at $\nu =0.5\,{\rm
GeV}$) lie within the range obtained from other estimates \cite{IS,WPC,BK}
and differ somewhat from the value $\chi \simeq -5.7\,{\rm GeV}^{-2}$ found
in the two- and three-pole models of Ref. \cite{BK}. 

In conclusion, we have recovered a third reliable sum rule for the nucleon
magnetic moments. In contrast to the other two, it receives previously
neglected direct-instanton contributions which arise from the interplay with
long-wavelength vacuum fields. Our new sum rule is built on the chirally-odd
amplitude $\Pi _{3}$ of the nucleon correlator in an electromagnetic
background field and found to be at least as stable as the other two
(although it had previously been regarded as flawed). The new sum rule adds
to the predictive power of the background-field sum rules and strengthens
their mutual consistency.

Moreover, our results reinforce a systematic pattern which emerged from the
study of direct-instanton effects in the pion \cite{for95}, nucleon \cite
{FB,for97} and glueball \cite{for200} channels: those sum rules which worked
satisfactorily without instanton corrections receive little or no direct
instanton contributions while previously less reliable or completely
unstable sum rules are stabilized by large instanton contributions. This
pattern points not only towards the importance of direct instantons in
particular sum rules, but also supports the adequacy of their semiclassical
implementation into the OPE. Our results show that these conclusions
continue to hold in the presence of a ``magnetized'' vacuum.

\section*{Acknowledgments}

I would like to thank the organizers for this interesting and
stimulating meeting, and for choosing such a beautiful location. I would
also like to thank Mountaga Aw and Manoj Banerjee for their collaboration on
the work presented here, which was supported by the Deutsche
Forschungsgemeinschaft under habilitation grant Fo 156/2-1 and by the U.S.
Dept. of Energy under grant number DE-FG02-93ER-40762.

\end{document}